\documentstyle[12pt]{article}

\begin{document}

\noindent
\hfill MSUCL-898

\noindent
\hfill December 1994

\setlength{\oddsidemargin}{0.6in}
\setlength{\evensidemargin}{0.6in}
\setlength{\textwidth}{6.5in}
\setlength{\textheight}{9in}
\setlength{\parskip}{0.05in}
\setlength{\baselineskip}{0.2in}
\vspace{36pt}

\newpage
\vspace{5pt}
\centerline{
\large{\mbox{\boldmath $\rho$},
\mbox{\boldmath $\omega$},
\mbox{\boldmath $\phi$}}\,\Large{\bf -Nucleon Scattering Lengths}}
\vspace{10pt}
\centerline{\Large{\bf from QCD Sum Rules}}
\par
\par\bigskip
\par\bigskip
\vspace{0.7cm}

\centerline{Yuji Koike
\footnote{Present address: Department of Physics,
Niigata University, Ikarashi, Niigata 950-21, Japan}}
\centerline{\em National Superconducting Cyclotron Laboratory,
Michigan State University}
\centerline{\em East Lansing, MI 48824-1321, USA}
\vspace{1in}

\centerline{\bf Abstract}
The QCD sum rule method is applied to derive a formula for
the $\rho$, $\omega$, $\phi$ meson-nucleon spin-isospin-averaged
scattering lengths $a_{\rho,\omega,\phi}$.
We found that the crucial matrix elements are
$\langle\bar{q}\gamma_\mu D_\nu q\rangle_N$ ($q=u,d$)
(twist-2 nucleon matrix element)
for $a_{\rho,\omega}$
and $m_s\langle\bar{s}s\rangle_N$ for $a_\phi$, and
obtained $a_\rho =0.14\pm 0.07$ fm,
$a_\omega =0.11\pm 0.06$ fm and
$a_\phi =0.035\pm 0.020$ fm.
These small numbers originate from a common factor
$1/(m_N+m_{\rho,\omega,\phi})$.  Our result
suggests a slight increase
($< 60$ MeV for $\rho$, $\omega$, and $<15$ MeV for $\phi$) of
the effective mass of these vector mesons in the nuclear matter
(in the {\it dilute} nucleon gas approximation).
The origin of the discrepancy with the previous study
was clarified.

\vspace{1cm}


\newpage
\setcounter{equation}{0}
\renewcommand{\theequation}{\arabic{equation}}

The operator product expansion (OPE) provides us with a
convenient tool
to decompose a variety of correlation functions
into the perturbatively calculable c-number coefficients and the
nonperturbative
matrix elements.
In its application to the QCD sum rules (QSR)\,\cite{SVZ,RRY},
the OPE supplies
an expression for the resonance parameters in terms of the
vacuum condensates
representing the nonperturbative dynamics in the correlators.
In the application to the deep inelastic
scattering (DIS)\,\cite{Muta},
the OPE isolates
the quark-gluon distribution functions of the target from
the short distance
cross sections.  In this paper,
we investigate the vector meson
($\rho$, $\omega$, $\phi$)-nucleon scattering lengths
utilizing these two
aspects of the OPE.
These scattering lengths can be
measured through
the photo-production of these vector mesons.
Furthermore, they determine the mass shift
of the vector mesons in the dilute nuclear medium.
This will be discussed in the last part of this paper.
A similar idea was recently presented for
the nucleon-nucleon scattering
length in ref.\,\cite{KM}.

We start our discussion with the forward
scattering amplitude of the
vector current $J_\mu^V$ ($V=\rho,\omega,\phi$) off the
nucleon target with the
four momentum $p=(p^0,
\mbox{\boldmath $p$})$ and the polarization $s$:
\begin{eqnarray}
T_{\mu\nu}(\omega,
\mbox{\boldmath $q$})=i\int\,d^4x\,e^{iq\cdot x}\langle ps|
T\left( J^V_\mu(x) J^V_\nu(0) \right) |ps\rangle,
\label{eq1}
\end{eqnarray}
where $q=(\omega,\mbox{\boldmath $q$})$ is
the four-momentum carried by $J_\mu^V$
and the nucleon state is normalized covariantly
as $\langle p |p'\rangle = (2\pi)^3 2p^0
\delta(\mbox{\boldmath $p$}-\mbox{\boldmath $p'$}) $.  We set
$p=(m_N,{\bf 0})$ throughout this work and
suppress the explicit dependence on $p$ and $s$.
The vector current $J^V_\mu$ is defined as
$J_\mu^{\rho,\omega}(x)
=(1/2)(\bar{u}\gamma_\mu u(x)\mp\bar{d}\gamma_\mu d(x))$,
$J_\mu^{\phi}(x)= \bar{s}\gamma_\mu s(x)$.
Near the pole position of the vector meson, $T_{\mu\nu}$ can
be associated with the $T$-matrix
for the forward $V-N$ helicity amplitude,
${\cal T}_{hH,h'H'}(\omega,\mbox{\boldmath $q$})$
($h(h')$ and $H(H')$ are
the helicities of the initial (final) vector meson
and the initial (final) nucleon,
respectively, and they take the values of
$h,h'=\pm 1,0$ and $H,H'=\pm 1/2$) as
\begin{eqnarray}
\epsilon^{(h)}_\mu (q) T_{\mu\nu}(\omega,\mbox{\boldmath $q$})
\epsilon^{(h')*}_\nu (q)
\simeq -{f_V^2m_V^4 \over (q^2-m_V^2+i\eta)^2 }
 {\cal T}_{hH,h'H'}(\omega,\mbox{\boldmath $q$}),
\label{eq3}
\end{eqnarray}
where we introduced the coupling $f_V$ and the mass $m_V$
of the vector
meson $V$ by the relation
$\langle 0|J_\mu^V|V^{(h)}(q)\rangle
= f_V m_V^2 \epsilon_\mu^{(h)}(q)$ with
the polarization vector $\epsilon_\mu^{(h)}(q)$ normalized as
$\sum_{\rm pol.}\epsilon_\mu^{(h)}(q)
\epsilon_\nu^{(h)}(q)=-g_{\mu\nu}
+q_\mu q_\nu/q^2$.
As is well known in DIS,
$T_{\mu\nu}$ can be decomposed into the four
scalar components respecting
the current conservation and the invariance
under parity and
time-reversal. (Two of them correspond to the spin-averaged
structure functions $W_1$ and $W_2$,
and the other two to the spin-dependent
ones $G_1$ and $G_2$.)  Correspondingly, there are 4 independent
helicity amplitudes for the vector current-nucleon scattering;
${\cal T}_{1{1\over 2},1{1\over 2}}$,
${\cal T}_{1{-1\over 2},1{-1\over 2}}$,
${\cal T}_{0{1\over 2},0{1\over 2}}$,
${\cal T}_{1{-1\over 2},0{1\over 2}}$, all the
rest being obtained
by time-reversal and parity from these four.
Since the information on $G_1$ and $G_2$
is still lacking, we shall focus on
the combination
$T=T_1+(1-(p\cdot q)^2/m_N^2q^2)T_2$
(${\rm Im}\,T_i\sim W_i,\ i=1,2$),
which projects the $V-N$ spin-averaged $T$-matrix,
${\cal T}(\omega,\mbox{\boldmath $q$})$.
In the low energy limit ($\mbox{\boldmath $q$}
 \rightarrow {\bf 0}$),
${\cal T}$ is reduced to the
$V-N$ spin-averaged scattering length $a_V=(1/3)(a_{1/2}
+2a_{3/2})$ ($a_{1/2}$ and $a_{3/2}$ are the
scattering lengths in the spin-1/2 and 3/2 channels,
respectively) as
${\cal T}(m_V,{\bf 0})=24\pi(m_N+m_V)a_V$\,\cite{Pil}.
A useful quantity for the dispersion analysis is the
retarded correlation
function defined as
\begin{eqnarray}
T^R_{\mu\nu}(\omega,
\mbox{\boldmath $q$})=i\int\,d^4x\,e^{iq\cdot x}
\theta(x^0) \langle ps|
\left[ J^V_\mu(x), J^V_\nu(0) \right] |ps\rangle
={1\over \pi}
\int_{-\infty}^{\infty}\,du\,{ {\rm Im}\,T^R_{\mu\nu}(u,
\mbox{\boldmath $q$})
\over  u-\omega -i\eta },
\label{eq5}
\end{eqnarray}
which is analytic in the upper half $\omega$-plane with
a fixed $\mbox{\boldmath $q$}$.
Noting the crossing symmetry,
the $V-N$ scattering
contribution to the spin-averaged spectral function
at $\mbox{\boldmath $q$} = {\bf 0}$
can be written as
\begin{eqnarray}
& &{1\over\pi}{\rm Im}\left[
T^R(\omega,{\bf 0}) \right]
=\theta(\omega){1\over\pi}{\rm Im}\left[
T(\omega,{\bf 0}) \right]
-\theta(-\omega)
{1\over\pi}{\rm Im}\left[
T(\omega,{\bf 0}) \right]
\nonumber\\
& &\quad =
-24\pi f_V^2m_V^4(m_N+m_V)a_V
\left[\theta(\omega)\delta'(\omega^2-m_V^2)
-\theta(-\omega)\delta'(\omega^2-m_V^2)\right]+({\rm S.\,P.}),
\label{eq6}
\end{eqnarray}
where $\delta'(x)$ is the first derivative of the
$\delta$-function (double pole term)
and (S.\,P.) denotes the simple pole term representing
the off-shell energy dependence of the $T$-matrix.
Equation (\ref{eq6}) can also be derived starting from the
spectral representation.
Using this form of the spectral function in eq.\,(\ref{eq5}),
and noting that the retarded
correlation
function $T^R_{\mu\nu}$ becomes
identical to the causal correlation function $T_{\mu\nu}$
in the deep Euclidean region $\omega^2 =-Q^2 \rightarrow -\infty$,
one gets
\begin{eqnarray}
T(\omega^2=-Q^2)
= -24\pi f_V^2 m_V^4(m_N+m_V)a_V
{1 \over (m_V^2 + Q^2)^2 }+{\cal R}_{os}(Q^2)+{\cal R}_c(Q^2),
\label{eq7}
\end{eqnarray}
where
we have used the fact that $T$ becomes a function of
$\omega^2$ in this limit.
In eq.\,(\ref{eq7}), we assumed the
spectral function is saturated by the $V-N$
scattering (with its off-shell effect) and the
"continuum" contribution\,\cite{aspt1}:
${\cal R}_{os}(Q^2)$ denotes the simple pole term
corresponding to the off-shell part of the $V-N$ $T$-matrix
($\sim 1/(m_V^2+Q^2)$) and ${\cal R}_{c}(Q^2)$
stands for the "continuum" contribution
with its threshold $S_0'$ ($\sim 1/(S_0'+Q^2)$)\,\cite{aspt2}.
The sum of the residues of ${\cal R}_{os}(Q^2)$ and
${\cal R}_{c}(Q^2)$ is constrained by the $1/Q^2$-term in the
OPE side of the correlator. (See eq.\,(\ref{eq9}) below.)

We now proceed to the OPE side of $T(\omega^2=-Q^2)$ (l.h.s. of
eq.\,(\ref{eq7})).
Unlike in DIS, our OPE is the short distance expansion
and hence all the
operators with the same dimension contribute in the same order
with respect to $1/Q^2$ ($=-1/\omega^2$ at
$\mbox{\boldmath $q$}={\bf 0})$.
The complete OPE expression
for $T(Q^2)$ including the operators up to dimension=6
is given in ref.\,\cite{HKL} in the context of the finite
temperature QSR.
For the $\rho$ and $\omega$ mesons,
it reads from eq.\,(2.13) of \cite{HKL} as ($-$ for $\rho$
and $+$ for $\omega$)
\begin{eqnarray}
T^{\rho,\omega}(Q^2) &=& {1\over 4}\left[
-{2m_q \over Q^2}\langle \bar{u}u+ \bar{d}d
\rangle_N - {1\over 6Q^2}
\langle{\alpha_s \over \pi} G^2
\rangle_N + {2\pi\alpha_s \over Q^4}
\left(\langle{\cal Q}^\mp_5 + {2\over 9}{\cal Q}^+
\rangle_N \right)
\right]\nonumber\\
& &-{m_N^2\over 2Q^2}A_2^{u+d} +{5m_N^4\over 6Q^4}A_4^{u+d}
-{m_N^2 \over 2Q^4}
\left( B_{1\mp}+{1\over 4}B_2 +{5\over 8}B_3
\right),
\label{eq9}
\end{eqnarray}
where $\langle\cdot\rangle_N$ denotes the spin-averaged
nucleon matrix element, and
${\cal Q}_5^\mp$ and ${\cal Q}^+$ are the scalar
four-quark operators
familiar in the QSR for the $\rho$ and $\omega$ mesons;
${\cal Q}_5^\mp = (\bar{u}\gamma_\mu\gamma_5\lambda^a u
\mp \bar{d}\gamma_\mu\gamma_5\lambda^a d)^2$,
${\cal Q}^+ = \left( \bar{u}\gamma_\mu\lambda^a u
+ \bar{d}\gamma_\mu\lambda^a d \right)
\sum_q^{u,d,s}\bar{q}\gamma_\mu\lambda^a q$.
In eq.\,(\ref{eq9}),
$A_n^{u+d} \equiv A_n^u+A_n^d\ (n=2,4)$ are related to the
twist-2
operators and are given as the $n$-th moment of the parton
distribution
function ($q=u,d,s$);
$\langle {\cal ST}\left( \bar{q}\gamma_{\mu_1}D_{\mu_2}
\cdot\cdot\cdot D_{\mu_n} q(\mu) \right)\rangle_N
=(-i)^{n-1} A_n^q(\mu) (p_{\mu_1}\cdot\cdot\cdot
p_{\mu_n} -{\rm traces})$,
$A^q_n(\mu) = 2\int_0^1
\,dx\,x^{n-1}\left(q(x,\mu)+(-1)^{n}\bar{q}(x,\mu)\right)$
with the renormalization scale $\mu$.
$B_i$ ($i=1\mp,2,3$)
are associated with the twist-4 matrix elements as
$\langle {\cal O}^i_{\mu\nu}(\mu) \rangle_N =
(p_\mu p_\nu -m_N^2g_{\mu\nu}/4)B_i(\mu)$ with
${\cal O}_{\mu\nu}^{1\mp} =
(g^2/4){\cal ST}\left((\bar{u}\gamma_\mu\gamma_5{\lambda^a}u
\mp \bar{d}\gamma_\mu\gamma_5{\lambda^a} d)(\mu\rightarrow
\nu)\right)$,
${\cal O}_{\mu\nu}^2 =
(g^2/4){\cal ST}\left((\bar{u}\gamma_\mu{\lambda^a}u
+ \bar{d}\gamma_\mu{\lambda^a}d)
\sum_{q}^{u,d,s}\bar{q}\gamma_\nu{\lambda^a}q\right)$,
${\cal O}_{\mu\nu}^3 =
ig{\cal ST}\left( \bar{u}\{D_\mu, ^*G_{\nu\lambda}\}
\gamma^\lambda\gamma_5 u
+ (u \rightarrow d) \right)$, where
the color matrix $\lambda^a$ is normalized as ${\rm Tr}
(\lambda^a\lambda^b)=2\delta^{ab}$ and
the symbol ${\cal ST}$ makes
the operators
symmetric and traceless with respect to the Lorentz indices.

To get an expression for the $V-N$ scattering length,
we first make a Borel
transform of eqs.\,(\ref{eq7}) and (\ref{eq9})
with respect to $Q^2$, and then eliminate the unknown parameter
which determines the ratio between the coefficient
of ${\cal R}_{os}(Q^2)$ and ${\cal R}_{c}(Q^2)$ ,
using the sum rule obtained by
taking the derivative with respect to the
Borel mass $M^2$.
We also eliminate the unknown coupling constant
$f_V$ by taking the ratio between the obtained sum rule and
the QSR
expression for the vector current correlators in the vacuum.
We thus get an expression
for the spin-averaged scattering length $a_V$ as
\begin{eqnarray}
a_{\rho,\omega} =
{\pi M^2 \over 3m_{\rho,\omega}^2(m_N+m_{\rho,\omega})}
{ r\beta/(\alpha M^2) + t\gamma/(\alpha M^4)
\over (1+{\alpha_s\over \pi})
(1-e^{-S_0/M^2})
+ b/M^4 -c/M^6 },
\label{eq13}
\end{eqnarray}
with
\begin{eqnarray}
r&=&m_N\Sigma_{\pi N} -
{2\over 27}m_0^2 + {m_N^2\over 2}A_2^{u+d},
\nonumber\\
t&=& -{112\pi\alpha_s \over 81}\left(
\langle\bar{u}u\rangle \langle\bar{u}u
\rangle_N + \langle\bar{d}d\rangle
\langle\bar{d}d\rangle_N \right)
-{5\over 6} m_N^4 A_4^{u+d} + {m_N^2\over 2}
\left( B_{1\mp} + {1\over 4}B_2 + {5\over 8}B_3 \right),
\nonumber\\
b&=&4\pi^2m_q\langle \bar{u}u + \bar{d}d \rangle
 + {\pi^2 \over 3}
\langle {\alpha_s \over \pi} G^2 \rangle,
\nonumber\\
c&=& {448\pi^3\alpha_s \over 81} \langle \bar{u} u
\rangle^2,
\nonumber
\end{eqnarray}
where $\langle\cdot\rangle$ denotes the vacuum condensate and
$S_0$ is the continuum threshold in the vacuum sum rule.
The factors $\alpha$, $\beta$ and $\gamma$ appeared through the
process of eliminating the parameter which determines
the ratio between the residues of ${\cal R}_{os}(Q^2)$
and ${\cal R}_{c}(Q^2)$, and they are defined as
$\alpha=1-e^{-(S_0'-m^2)/M^2}(1+{S_0'-m^2 \over M^2})$,
$\beta={m^2 \over M^2}+{S_0'-m^2 \over M^2}e^{-S_0'/M^2}
-{S_0'\over M^2}e^{-(S_0'-m^2)/M^2}$,
$\gamma=1+{m^2 \over M^2}
-(1+{S_0'\over M^2})e^{-(S_0'-m^2)/M^2}$
with $m=m_{\rho,\omega}$.
If we ignore ${\cal R}_{c}(Q^2)$ from the beginning,
the corresponding formula
is obtained by the replacement; $\alpha \rightarrow 1$,
$\beta \rightarrow 1-e^{- m^2/M^2}$,
$\gamma \rightarrow 1$.
In eq.\,(\ref{eq13}), we have used the following relations
for the matrix elements
as has been used in the study of QCD sum rules
in the nuclear matter
\,\cite{DL,HL}:
(i) $\pi N$ $\sigma$-term $\Sigma_{\pi N}$ is introduced
through the relation
$m_q\langle\bar{u}u+\bar{d}d\rangle_N=2m_N\Sigma_{\pi N}$.
(ii) The nucleon mass in the chiral limit, $m_0$,
is introduced in favor of
$\langle{\alpha_s \over \pi}G^2\rangle_N$
through the QCD trace
anomaly: $\langle{\alpha_s \over \pi}G^2\rangle_N=-(16/9)m_0^2$.
(iii) Factorization
is assumed for the
vacuum four-quark condensates,
$\langle {\cal Q}_5^\mp \rangle$
and $\langle {\cal Q}^+ \rangle$, as is usually
adopted in the literature\,\cite{SVZ,RRY}.
(iv) Factorization is also employed to estimate
the nucleon matrix elements
of the scalar four-quark operators
$\langle {\cal Q}_5^\mp \rangle_N$ and $\langle {\cal Q}^+
\rangle_N$
after making the Fierz transform\,\cite{HL},
i.e. $\langle (\bar{q}\Gamma\lambda q)^2
\rangle_N \rightarrow \langle
(\bar{q}q)^2 \rangle_N
\simeq 2\langle\bar{q}q\rangle\langle\bar{q}q\rangle_N$.

By repeating the same steps as above for $J^\phi_\mu$, one gets
the spin-averaged $\phi-N$ scattering length as
\begin{eqnarray}
a_{\phi} = {\pi M^2 \over 3m_{\phi}^2(m_N+m_{\phi})}
{ r_s\beta/(\alpha M^2)
+ t_s\gamma/(\alpha M^4)
\over (1+{\alpha_s\over \pi})(1-e^{-S_0/M^2})
-6m_s^2/M^2 + b_s/M^4 -c_s/M^6 },
\label{eq14}
\end{eqnarray}
with
\begin{eqnarray}
r_s&=&m_s\langle\bar{s}s\rangle_N
- {2\over 27}m_0^2 + m_N^2 A_2^{s},
\nonumber\\
t_s&=& -{224\pi\alpha_s \over 81}
\langle\bar{s}s\rangle \langle\bar{s}s\rangle_N
-{5\over 3} m_N^4 A_4^{s} + m_N^2
\left( B_1^s + {1\over 4}B_2^s + {5\over 8}B_3^s \right),
\nonumber\\
b_s&=&8\pi^2m_s\langle \bar{s}s \rangle + {\pi^2 \over 3}
\langle {\alpha_s \over \pi} G^2 \rangle,\nonumber\\
c_s&=& {448\pi^3\alpha_s
\over 81} \langle \bar{s} s\rangle^2,\nonumber
\end{eqnarray}
where
the strange twist-4 matrix elements
$B_i^s$ ($i=1-3$) are defined similarly to
the case of the $\rho$ and $\omega$ mesons.
For the vacuum condensates and the quark masses
in eqs.\,(\ref{eq13}) and (\ref{eq14}), we use the
standard values at the
renormalization scale $\mu=1$ GeV\,\cite{RRY}:
$\alpha_s=0.36$, $m_q=7$ MeV,
$m_s=110$ MeV,
$\langle\bar{u}u\rangle=\langle\bar{d}d\rangle=
(-0.28\,{\rm GeV})^3$, $\langle\bar{s}s\rangle
=0.8\langle\bar{u}u\rangle$.
With these vacuum condensates and the continuum threshold
$S_0=1.75$ GeV$^2$ for $\rho$, $\omega$ and $S_0=2.0$ GeV$^2$ for
$\phi$,
the experimental values
for $m_{\rho,\omega,\phi}$ are well reproduced.
We thus fixed $S_0$ at these values and use $m_{\rho,
\omega}=770$ MeV and $m_\phi=1020$ MeV in eqs.\,(\ref{eq13})
and (\ref{eq14}).
As a measure of the strangeness content of the nucleon,
we introduce the parameter
$y=2\langle\bar{s}s\rangle_N/(\langle\bar{u}u\rangle_N
+\langle\bar{d}d\rangle_N)$ and write
$\langle\bar{s}s\rangle_N = y m_N\Sigma_{\pi N}/m_q$.
For the nucleon matrix elements we use
$\Sigma_{\pi N}=45 \pm 7$ MeV,
$y=0.2$ and $m_0=830$ MeV obtained by the chiral
perturbation theory\,\cite{GLS}.
Since we ignored the twist-2 gluon operators in
eqs.\,(\ref{eq13}) and (\ref{eq14}),
we consistently use the
leading order (LO) parton distribution functions
of Gl\"uck, Reya and Vogt\,\cite{GRV} to determine $A_i^{u+d}$
and $A_i^s$ ($i=2,4$).  It gives
$A_2^{u+d}=0.90$,
$A_4^{u+d}=0.12$,
$A_2^{s}=0.05$ and $A_4^s=0.002$ at $\mu^2=1$ GeV$^2$.
For the twist-4 matrix elements
$B_i$ and
$B_i^s$, we use our recent result
\cite{CHKL} extracted from the newest DIS
data at CERN and SLAC.  It is based on the SU(2) flavor
symmetry (i.e. $B_i^s=0$ ($i=1-3$)) and a mild assumption on the
matrix elements invoked by the flavor structure
of the twist-4 operators.
Both for the proton and the neutron,
it gives $B_{1\mp} + B_2/4 +5B_3/8=-0.24\pm0.15\
(-0.41\pm 0.23)$ GeV$^2$ for
the $\rho$
($\omega$) meson at $\mu^2=5$ GeV$^2$\,\cite{aspt3}.
[Note that our $\rho^0-N$ ($N$ can be either proton or neutron)
scattering length correspond to the isospin-spin averaged one.]

Using these numbers for the matrix elements,
the Borel curves
for the $\rho,\omega,\phi$-nucleon
scattering lengths $a_{\rho,\omega,\phi}$
(eqs.\,(\ref{eq13})
and (\ref{eq14})) are shown
in Fig.\,1.
We determined the values
of $S_0'$ in order to minimize the slope of
the curves at $0.8 < M^2 < 1.3$ GeV$^2$.  They are 3.32 GeV$^2$
for $\rho$, 3.29 GeV$^2$ for $\omega$
and 4.40 GeV$^2$ for $\phi$.
With the above parameters, $r$ in eq.\,(\ref{eq13}) reads
$r=0.04-0.05+0.40=0.39$ GeV$^2$
from the first to the third terms.
Thus $r$ is totally dominated
by the twist-2 nucleon matrix element $A_2^{u+d}$ and the
cancelling contribution from the first and the second terms
makes the ambiguity in
$\Sigma_{\pi N}$ and $m_0$ less important.
The $t$-term in eq.\,(\ref{eq13}) reads
$t=0.42-0.08-0.11\pm 0.07\,(-0.18\pm 0.10)
=0.23\pm 0.07\,(0.16\pm 0.10)$ GeV$^4$ for $\rho$
($\omega$), which shows the contribution from the twist-4
matrix elements is sizable.
To get an insight on the sensitivity of the results to the
variation of $t$,
we also showed $a_{\rho,\omega}$
without the twist-4 matrix elements in $t$
with $S_0'=3.35$ GeV$^2$.
One sees that the inclusion of $B_i$ reduces
the $a_{\rho,\omega}$
by about
20 \% (30 \%) for $\rho$ ($\omega$).
With the uncertainty in $B_i$ in mind,
we assign error bars as $a_\rho=0.14\pm 0.07$ fm and
$a_\omega=0.11\pm 0.06$ fm,
taking the values for $a_{\rho,\omega,\phi}$
around $M^2=1$ GeV$^2$.
For the case of $a_\phi$,
the value of $m_s\langle\bar{s}s\rangle_N$ governs
the whole result because of large $m_s$,
i.e. $r_s=0.13-0.05+0.04=0.12$ GeV$^2$ and
$t_s=0.066-0.003+({\rm twist-4}\equiv 0)=0.063$ GeV$^4$ from
the first to
the third terms in $r_s$ and $t_s$.
Due to the uncertainty in $m_s\langle\bar{s}s\rangle_N$,
we read from Fig.\,1 $a_\phi=0.035 \pm 0.020$ fm.
Some phenomenological
analyses on the nucleon form factor\,\cite{GH}
and the nuclear force\,\cite{NRS}
suggest quite a large OZI violating $\phi NN$ coupling constant
$g_{\phi NN}
/g_{\omega NN} \sim 0.4$.  Equation (\ref{eq14})
supplies a neat expression for the $\phi N \rightarrow \phi N$
interaction strength in
terms of the strangeness content of the nucleon,
showing the importance of $m_s\langle\bar{s}s\rangle_N$
rather than
$\langle\bar{s}\gamma_\mu D_\nu s \rangle_N$.

If we calculate the scattering lengths
without ${\cal R}_{c}(Q^2)$ in eq.\,(\ref{eq7}), we get
even smaller numbers for the scattering lengths:
$a_{\rho} \sim 0.1$ fm,
$a_{\omega} \sim 0.08$ fm and $a_{\phi} \sim 0.01$ fm
around $M = 1$ GeV.   This way, the actual numbers for
$a_{\rho,\omega,\phi}$ depends on the assumption made
in the spectral function, although their typical
order of magnitude does not change.

One might be surprised by the smallness of these
scattering lengths compared with a typical
hadronic size ($\sim 1$ fm).
 From eqs.\,(\ref{eq13})
and (\ref{eq14}), one sees
$a_V \sim 1/(m_N + m_V)$, since $r$ and $r_s$ are
dominated by the third and the first terms, respectively.
If one applies the present method to the axial vector
correlator, one can easily
get the pion-nucleon scattering length
in the isospin symmetric channel
as $a_{\pi N} \propto m_N \Sigma_{\pi N} /(f_\pi^2 (m_N+m_\pi))$,
which is the same result as that of the current algebra\,\cite{KMN}.
[In the chiral limit,
$a_{\pi N}=0$, since $\Sigma_{\pi N} =0$.]
Therefore it is interesting to observe that our
method of deriving the
vector meson-nucleon scattering length is a generalization
of the current algebra technique for the
pion-nucleon scattering
length.
For the vector meson case,
the common factor $1/(m_N+m_V)$ makes $a_V$ small.
We believe this smallness
of the $V-N$ scattering lengths somehow sketches the real
situation, although the actual numbers for $a_V$ are
not trustable
because of the simplified form for the spectral
function in our calculation as was noted before.
A model calculation of the $\rho -N$
scattering amplitude based on an effective
hadronic lagrangian
suggests a similar
small number for $a_\rho$\,\cite{HFN}.

Let us finally discuss the mass shift of the vector mesons
in the nuclear
medium using the result for the scattering lengths here.
In the {\it dilute} nucleon gas approximation,
the $V$-current correlator
in the nuclear medium can be written as
\begin{eqnarray}
\Pi^{N.M.}_{\mu\nu}(\omega,\mbox{\boldmath $q$})=
i\int\,d^4x\,e^{iq\cdot x}\langle
T\left( J^V_\mu(x) J^V_\nu(0) \right) \rangle
+\sum_{\rm pol.}\int^{p_f} {d^3p \over (2\pi)^3 2p^0}
T_{\mu\nu}(\omega,\mbox{\boldmath $q$}).
\label{eq15}
\end{eqnarray}
By ignoring the Fermi motion of the nucleon,
$\Pi^{N.M.}_{\mu\nu}(\omega,\mbox{\boldmath $q$}
={\bf 0})$ can be
approximated near the pole position as
\begin{eqnarray}
\Pi^{N.M.}_{\mu\nu}(\omega,{\bf 0})&\simeq&
f_V^2 m_V^4 \left( g_{\mu\nu}
- {q_\mu q_\nu \over \omega^2}\right)
\left( {1 + O(\rho_N)
\over \omega^2 -m_V^2} + \Delta m_V^2 {1 \over
(\omega^2 - m_V^2 )^2} \right)\nonumber\\
&\sim& {1+O(\rho_N) \over \omega^2 -m_V^2 -\Delta m_V^2}
+O(\rho_N^2),
\label{eq16}
\end{eqnarray}
where
$\Delta m_V^2 = 12\pi a_V \rho_N (m_N +m_V)/m_N$
with the nucleon density $\rho_N$.  From this relation,
$\Delta m_V^2$ can be viewed as a shift of $m_V^2$
in the nuclear
medium\,\cite{aspt4}.
Our values for $a_{\rho,\omega,\phi}$ suggest
that the effective
mass for the vector mesons
increases by about $27-57$ MeV for $\rho$,
$20-48$ MeV
for $\omega$ and $5-13$
MeV for $\phi$ at the nuclear matter density
$\rho_N =0.17$ fm$^{-3}$\,\cite{Chin}.
[Note that the validity of the mass shift
discussed here hinges on the
assumption that the off-shell energy dependence
and the momentum dependence of the $V-N$ scattering
amplitude is weak within the range of the
nucleon's Fermi momentum.]

The authors of ref.\,\cite{HL} applied the QSR
method to study mass shifts of the
$\rho$, $\omega$ and $\phi$ mesons in the nuclear medium.
Although their approximation in the OPE side of the correlation
functions is essentially the same as ours,
eq.\,(\ref{eq15}), they found a
serious {\it decrease} of these vector meson masses
in the nuclear matter.
Here we clarify the origin of this discrepancy.
In the recent literature of the QSR method
in the nuclear medium for baryons and mesons\,\cite{DL,HL},
the common starting point
is that the density dependence of correlation
functions is ascribed to the density dependent condensates:
\begin{eqnarray}
\Pi^{N.M.}(q, \rho_N) = \sum_i C_i(q, \mu)
\langle {\cal O}_i(\mu) \rangle_{\rho_N},
\label{eq17}
\end{eqnarray}
where $C_i(q, \mu)$ and ${\cal O}_i(\mu)$ are the
Wilson coefficient
and a local operator, respectively, and we suppressed all the
spinor and Lorentz indices.  In the dilute nuclear medium,
$\langle {\cal O}_i(\mu) \rangle_{\rho_N}$ has been approximated
as
\begin{eqnarray}
\langle {\cal O}_i(\mu) \rangle_{\rho_N}
&=&\langle {\cal O}_i(\mu) \rangle +
\sum_{\rm pol.}\int^{p_f} {d^3p \over (2\pi)^3 2p^0}
\langle ps|{\cal O}_i(\mu)|ps\rangle\nonumber\\[4pt]
&=&\langle {\cal O}_i(\mu) \rangle + {\rho_N \over 2m_N}
\langle {\cal O}_i(\mu) \rangle_N + o(\rho_N).
\label{eq18}
\end{eqnarray}
As is easily seen by inserting eq.\,(\ref{eq18}) into
eq.\,(\ref{eq17}), the
approximation for the condensate eq.\,(\ref{eq17})
is equivalent to
the approximation eq.\,(\ref{eq15}) for the correlator.
Therefore
the density dependent part of the correlator has to
be analyzed
from the point of view of the forward
current-nucleon scattering amplitude as was done in this paper.
In this approximation, what is relevant for the mass shift is
the double pole structure at the pole position
appearing in the forward amplitude.

To understand the difference between our result and
the one in \cite{HL},
it is convenient to recall the QSR for the vector
meson in the vacuum.
In the vacuum, the vector current correlator can be written as
$\Pi_{\mu\nu}(q)=
(q_\mu q_\nu-g_{\mu\nu}q^2)\Pi(q^2)$.  As long as one uses
a spectral function with a single
narrow resonance and the continuum, both $\Pi(q^2)$
and $q^2\Pi(q^2)$
can be used for the QSR analysis.  The formula for
the vector meson mass
in the Borel sum rule
obtained by using these two sum rules
are, respectively
\begin{eqnarray}
{m_V^2 \over M^2} &=& {
(1+ {\alpha_s \over \pi} )[1-(1+S_0/M^2)e^{-S_0/M^2}]
-D_4/M^4 +D_6/M^6
\over
(1+ {\alpha_s \over \pi} )[1-e^{-S_0/M^2}]
+D_4/M^4 -D_6/2M^6
},
\label{eq19}\\[8pt]
{m_V^2 \over M^2} &=& {
2(1+ {\alpha_s \over \pi} )[1-(1+S_0/M^2+S_0^2/2M^4)
e^{-S_0/M^2}] -D_6/M^6
\over
(1+ {\alpha_s \over \pi} )
[1-(1+S_0/M^2)e^{-S_0/M^2}] -D_4/M^4 +D_6/M^6
},
\label{eq20}
\end{eqnarray}
where $D_4$ and $D_6$ are the sums of the dim.=4
and dim.=6 condensates,
respectively.  Both sum rules give right
amount for $m_V$.  However,
one should note that the sign of the power correction
due to the condensates is opposite between the two sum rules.
Thus the shift of the condensates due to the
medium effects (the second term
of eq.\,(\ref{eq15}) or eq.\,(\ref{eq18}))
is expected to cause opposite physical effects
depending on which sum rule one uses.
The authors of \cite{HL} analyzed
$\Pi^{N.M.}(\omega,{\bf 0})=
\Pi^{N.M.\mu}_\mu(\omega,{\bf 0})/(-3\omega^2)$,
with
a simple pole ansatz for the vector meson
(together with a scattering term)
in the spectral function, which picks up the same
effect of the shift of the condensate as eq.\,(\ref{eq19}).
On the other hand,
if we recognize that the second term in eq.\,(\ref{eq15})
is associated with the $V-N$ forward amplitude through
eq.\,(\ref{eq3}), we can easily see that it is
$\omega^2 \Pi^{N.M.}(\omega,{\bf 0})$ which has to be analyzed
with a simple pole ansatz in the order $O(\rho_N)$
as is shown in eq.\,(\ref{eq16}).
In this case, the vector meson mass
receives the effect of the change
of the condensate as is expected from
the formula
eq.\,(\ref{eq20}).
In $\Pi^{N.M.}(\omega,{\bf 0})$,
the density dependent part appears as a form of
$(\rho_N/2m_N)T(\omega,{\bf 0})/\omega^2$,
which brings a factor
$-\Delta m_V^2 (m_V^2/Q^2)$ ($Q^2=-\omega^2 >0$)
instead of $\Delta m_V^2$
in the first line of eq.\,(\ref{eq16}).  In this case,
due to the additional factor $1/Q^2$,
the double pole term
can not be incorporated into the mass shift.
Thus the use of eq.\,(\ref{eq19}) in the nuclear medium
is simply wrong!
Therefore an inadequate form of the spectral
function in \cite{HL} led to
a fictitious ``negative" mass shift.
Since the second term in eq.\,(\ref{eq15})
has an unique relation with the
$V-N$ $T$-matrix around the
pole position as is shown in eq.\,(\ref{eq3}),
we believe the mass shift of the vector mesons
in the nuclear medium in the context of the QSR should
be understood as presented in this work.

Finally, we wish to make a comment on the speculation
on the in-medium behavior of the hadron masses existing
in the literature.  From the finite energy sum rule analysis,
one gets the $\rho$ meson mass as $m_\rho \propto
|\langle \bar{\psi}\psi \rangle |^{1/3}$ {\it in the vacuum}.
Since $|\langle \bar{\psi}\psi \rangle |$
decreases in the nuclear medium
according to the formula eq.\,(\ref{eq18}),
one might naively expect $m_\rho$ would also
decrease\,\cite{Bro}.
We have illustrated, however, that a consistent organization
of the QCD sum rule does not predict such a behavior.
It would rather
support (within our crude approximation)
another naive expectation that
a tendency of $\rho-A_1$ degeneracy might occur in the
nuclear medium,
since the application of our method to the $A_1$ meson
gives decreasing $m_{A_1}$.   We also remind
the readers that (1)
$|\langle \bar{\psi}\psi \rangle_{T} |$
decreases as the temperature ($T$) goes up, while all hadron
masses stay
constant in the order $O(T^2)$\,\cite{EI}, (2) a
consistent organization of the
sum rule at finite temperature certainly gives the same behavior
\,\cite{aspt4}
unlike the above naive expectation, and (3) this
is because the pion-hadron
scattering length is zero in the chiral limit.

In conclusion, we have derived the $\rho$, $\omega$,
$\phi$ -nucleon spin-isospin averaged scattering lengths
$a_{\rho,\omega,\phi}$
from QCD sum rules.
We obtained very small
positive numbers (corresponding to
the repulsive interaction) for $a_{\rho,\omega,\phi}$,
$a_{\rho,\omega,\phi} \sim 1/(m_N+m_{\rho,\omega,\phi})$,
although the actual numbers depend on
the factorization assumption for the nucleon matrix element
of the scalar
four-quark operator as well as the simplified form for the
spectral function.
This result suggests slight increase of these vector mesons
in the nuclear medium, which is contradictory
to the previous result by Hatsuda and Lee\,\cite{HL}.
We have clarified the origin of this discrepancy
and pointed out the problem of the analysis in
ref.\,\cite{HL}.
The detail of the calculation will be published elsewhere.

\vspace{1.0cm}
\centerline{\bf Acknowledgement}
\noindent
I would like to thank T.\,Hatsuda,
S.\,H.\,Lee, P.\,Danielewicz,
F.\,Lenz, O.\,Morimatsu, E.\,V.\,Shuryak and K.\,Yazaki for
many valuable discussions and comments.
This work is supported in part
by the US National Science Foundation
under grant PHY-9017077.

\newpage

\centerline{\bf Figure Captions}
\vskip 15pt
\begin{description}

\item[Fig. 1] The Borel curves for the
$\rho$, $\omega$, $\phi$-nucleon scattering
lengths.
The dashed line denotes the one for $\rho$ and
$\omega$ without the twist-4 matrix elements in eq.\,(\ref{eq13}).

\end{description}

\begin{thebibliography}{99}
\bibitem{aspt1}
There could be other contributions to the spectral function
at $\omega < m_V$,
such as $\rho+N \rightarrow \pi +N$ and $\rho+N
\rightarrow \pi +\Delta$
for the case of $J^\rho_\mu$-correlator, for example.
Here we assume these continuum contributions
are included in ${\cal R}_{os}(Q^2)$.
\bibitem{aspt2}
This form of the "continuum" contribution can be
understood by integrating over the pole-position $m_V^2$
of the double-pole contribution starting at
some "threshold" $S_0'$.
The form of the off-shell effect
${\cal R}_{os}(Q^2)$
can be reinterpreted in the same way.
\bibitem{aspt3}
We ignore the correction
due to the $\mu^2$-evolution
down to $\mu^2\sim 1$ GeV$^2$,
since the twist-4 anomalous dimensions are not available in the
literature.  But the magnitude of this correction
should be at most 20 \%, as it is logarithmic.
\bibitem{KMN} In a recent paper, more complete analysis
on the $\pi-N$ scattering lengths was made in a similar
QCD sum rule method: Y. Kondo, O Morimatsu and Y. Nishino,
INS-Rep.-1059.
\end{thebibliography}
\end{document}